\documentstyle[amssymb,aps,twocolumn]{revtex}

\input  epsf
\def\bea{\begin{eqnarray}}
\def\eea{\end{eqnarray}}
\def\be{\begin{equation}}
\def\ee{\end{equation}}

\def\mysize{5cm}

\begin{document}

\title{Stretched exponential relaxation for growing interfaces in quenched 
disordered media}

\author{A. D\'{\i}az-S\'anchez$^1$, A.\ P\'erez-Garrido$^1$, A.\ Urbina$^2$, 
and J.D.\ Catal\'a$^1$}
\address{$^1$Departamento de F\'\i sica Aplicada,
Universidad Polit\'ecnica de Cartagena,
Campus Muralla del Mar, Cartagena,
E-30202 Murcia, Spain. }
\address{$^2$Departamento de Electr\'onica, Tecnolog\'{\i}a de Computadoras y
Proyectos, Universidad Polit\'ecnica de Cartagena,
Campus Muralla del Mar, Cartagena,
E-30202 Murcia, Spain. }

\date{\today}
\wideabs{%
\maketitle
\begin{abstract}
We study the relaxation for growing interfaces in quenched disordered
media. We use a directed percolation depinning model introduced
by Tang and Leschhorn for $1+1$-dimensions. We define the two-time 
autocorrelation function of the interface height $C(t',t)$ and its Fourier 
transform. These functions depend on the difference of times $t-t'$ for 
long enough times, this is the steady-state regime. We find a two-step 
relaxation decay in this regime. The long time tail can be fitted by a 
stretched exponential relaxation function. The relaxation time 
$\tau_\alpha$ is proportional to the characteristic distance of the 
clusters of pinning cells in the direction parallel to the interface 
and it diverges as a power law. The two-step relaxation is lost at a 
given wave length of the Fourier transform, which is proportional to 
the characteristic distance of the clusters of pinning cells in the 
direction perpendicular to the interface. The stretched exponential 
relaxation is caused by the existence of clusters of pinning cells and 
it is a direct consequence of the quenched noise.
\end{abstract}
}

\section{Introduction}

For decades the investigation of growing surfaces and interfaces has 
attracted much attention due to its importance in many fields, such as 
motion of liquids in porous media, growth of bacterial colonies, crystal 
growth, fronts of fire, etc.  In these problems we have a nonequilibrium 
interface. The $d$-dimensional interface described by a single 
valued-function $h({\bf x},t)$ evolves in a $d+1$-dimensional medium. 
The disorder affects the motion of the interface and leads to its roughness.
A phenomenological non-linear Langevin equation, the Kardar-Parisi-Zhang 
equation (KPZ) \cite{KP86}, and the directed percolation depinning (DPD) 
models \cite{TL92,BB92} have been used in order to study growing interfaces.
Two main kinds of disorder have been proposed in these models: the annealed
noise that depends only on time and the quenched disorder due to the 
inhomogeneity of the media, which does not depend on time. In the DPD models 
the disorder is quenched and they describe very well some experiments such 
as the growth of bacterial colonies and the motion of liquids in porous 
media. These models were proposed simultaneously by Tang and Leschhorn 
\cite{TL92} and Buldyrev {\sl et al.} \cite{BB92}. 

In many glassy systems a non-exponential relaxation is found when they are 
close to some temperature above to the static transition. As an example, in 
structural glasses a two-step relaxation decay is found near the so called 
``ideal glass transition'' \cite{G91}. The long relaxation step has the 
stretched exponential form 
\be
f(t)=f_0 \exp\left[-(t/\tau_\alpha)^\beta\right]\;,
\label{eq:1}
\ee
where $0<\beta <1$ does not depend on the temperature. There are two 
mechanisms driving non-exponential relaxation. In disordered systems such 
as spin glasses that behavior is caused by the existence of non-frustrated 
ferromagnetic-type clusters of interactions \cite{RS85} which is a direct 
consequence of the quenched disorder\cite{FC97}. Another mechanism in 
frustrated systems is based on the percolation transition of the 
Kasteleyn-Fortuin and Coniglio-Klein cluster\cite{FK72}, here disorder 
is not needed to obtain non-exponential relaxation \cite{FF99}. Recently, 
Colaiori and Moore \cite{CM01} have found a stretched exponential relaxation 
for the KPZ equation with annealed noise.

In this paper, we use the DPD model proposed by Tang and Leschhorn (TL) 
\cite{TL92} in order to investigate the relaxation of the two-time 
autocorrelation functions in quenched disordered media at the steady-state regime. 
We relate the relaxation properties to the clusters of pinning cells. The paper 
is organized as follows. In Sec. II we present the model and some properties of 
the clusters of pinning cells. In Sec. III the steady-state relaxation is studied. 
Finally, in Sec IV we present some conclusions.

\section{The model}

In the TL model for $1+1$-dimensions \cite{TL92}, the advance of the fluid 
through the media is modeled by a driving force $p$, while the disorder of 
the media, that brakes this advance, is represented by a quenched noise in 
the substratum. The interface grows in a square lattice of edge $L$  with
periodic boundary conditions. We assign a random pinning force $g({\bf r})$ 
uniformly distributed in the interval $[0,1]$ to every cell of the square 
lattice. For a given applied driving force $p>0$ , we can divide the cells 
into two groups: those with $g({\bf r}) \le p$ (free cells), and those with
$g({\bf r}) > p$ (pinning cells). Denoting by $q$ the density of pinning cells 
on the lattice, we have $q=1-p$ for $0< p<1$ and $q=0$ for $p\ge 1$. The 
interface is specified completely by a set of integer column heights $h_i$ 
($i=1,\dots ,L$). At $t = 0$ all columns are assumed to have the same height, 
which is zero. During growth, a column is selected at random, say column $i$, 
and its height is compared with those of the neighbor columns $(i - 1)$ and
$(i + 1)$. The growth event is defined as follows. If $h_i$ is greater than 
either $h_{i-1}$ or $h_{i+1}$ by two or more units, the height of the lower 
of the two columns $(i-1)$ and $(i+1)$ is incremented in one (in case of the 
two being equal, one of them is chosen with equal probability). In the 
opposite case, $h_i < \min (h_{i-1},h_{i+1}) +2$, the column $i$ advances 
by one unit provided that the cell to be occupied is a free cell. Otherwise
no growth takes place. In this model, the time unit is defined as
one growth  attempt. In numerical simulations at each growth
attempt the time $t$ is increased by $\delta t$, where $\delta t =
1 / L$. Thus, after $L$ growth attempts the time is
increased in one unit. In our simulations we use $L=10000$ and take
the averages over 100 different realizations of quenched noise. 

\subsection{Clusters of pinning cells}

As it has been shown in Ref. \cite{TL92}, this model has a depinning transition 
at a driving force $p_c=0.461$. For driving forces below the critical one $p_c$, 
the advance of the interface is halted (pinning phase), 
while above this driving force the interface moves without stopping (moving phase). 
At the transition, the characteristic length $\xi$ of the pinned regions diverges.
A directed percolation cluster of pinning cells which extends over the whole system 
appears in the pinning phase. In the moving phase, a typical connected
cluster of pinned cells extends over a distance of the order of $\xi_\parallel$ 
in the direction parallel to the interface and a distance of the order of 
$\xi_\perp$ in the direction perpendicular to the interface. On both sides of 
the percolation transition, the two lengths have a power-law behavior 
$\xi_\parallel\sim |p-p_c|^{-\nu_\parallel}$ and 
$\xi_\perp\sim |p-p_c|^{-\nu_\perp}$, with 
$\nu_\parallel=1.733\pm 0.001$ and $\nu_\perp=1.097\pm0.001$. 
$\xi_\perp$ sets a characteristic scale
for the height while $\xi_\parallel$ sets characteristic scales for both the
distance parallel to the interface and the time. For the mean interface height
the scaling form $H(t)\approx \xi_\perp \Phi (t/\xi_\parallel)$ is obtained,
denoting $\Phi$ a scaling function which is different for the two phases. In the
moving phase there is a crossover from a power-law growth 
$H(t)\sim t^{\nu_\perp/\nu_\parallel}$ at $t\ll\xi_\parallel$ to
a linear behavior $H(t)=vt$ at $t\gg\xi_\parallel$. The steady-state velocity
can be expressed as $v(p)\sim(p-p_c)^{\nu_\parallel-\nu_\perp}$.

\section{Steady-state relaxation}

We define the two-time autocorrelation function of the surface height as 
\be
C(t',t)=\frac1L \sum_j \delta \left[ h_j(t')-h_j(t)\right]\;,
\ee
where $\delta\left[ x\right]$ is the delta function.
Its Fourier transform is 
\be
C_k(t',t)=\frac1L\sum_j e^{-i\left[h_j(t')-h_j(t)\right]k}\;,
\ee
where $k$ is the wave number.

\vglue 0.5cm

\begin{figure}
\epsfysize=\mysize
\begin{center}
\epsfbox{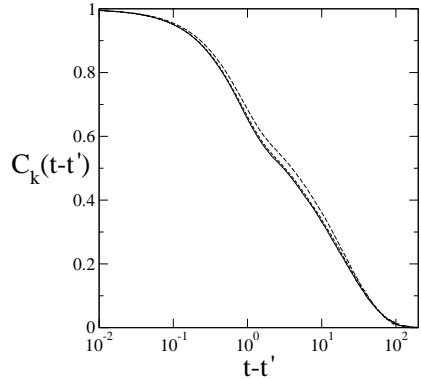}
\end{center}
\caption{
$C_k(t-t')$ for $p=0.5$ and $k=\pi$ and for different initial times 
$t'=10^2$ (long dashed line), $10^3$ (dashed line), $10^4$ (dotted line),
and $10^5$ (solid line). Dotted and solid lines overlap for any $t-t'$.
}
\label{fig:1}
\end{figure}

For long enough times, $t'\gg \xi_\parallel$, these functions depend only on the 
difference of times $t-t'$, this is the steady-state regime where $H(t)=vt$.
This regime is reached at longer times when we approach to the critical driving 
force $p_c$. 
Fig. \ref{fig:1} shows $C_k(t-t')$ for $p=0.5$ and $k=\pi$ and for different 
initial times $t'=10^2,\ 10^3,\ 10^4,$ and $10^5$. As we see $C_k(t-t')$ is 
independent of $t'$ when $t'\ge 10^4$ for this value of $p$.

\vglue 0.5cm

\begin{figure}
\epsfysize=\mysize
\begin{center}
\epsfbox{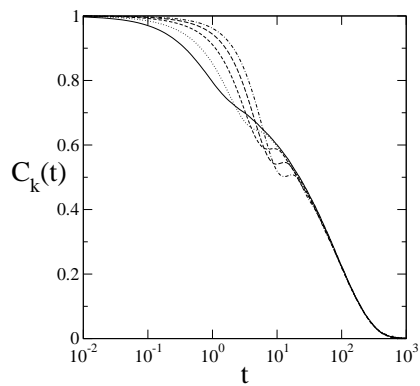}
\end{center}
\caption{ 
$C_k(t)$ in the steady-state regime for $p=0.5$ and $k=\pi$ (solid line),
$\pi/2$ (dotted line), $\pi/3$ (dashed line), $\pi/4$ (long dashed line), and 
$\pi/5$ (dot-dashed line).
}
\label{fig:2}
\end{figure}

\begin{figure}
\epsfysize=\mysize
\begin{center}
\epsfbox{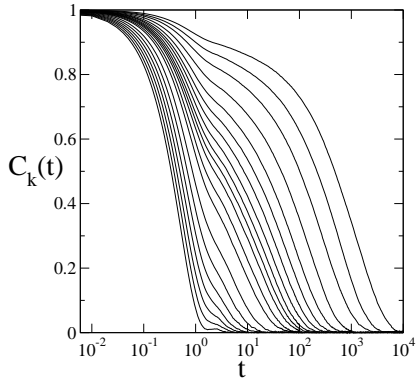}
\end{center}
\caption{
$C_k(t)$ in the steady-state regime for $k=\pi$ and $p=0.95$, 0.9, 0.85, 0.8, 0.75,
0.7, 0.65, 0.6, 0.68, 0.56, 0.55, 0.54, 0.53, 0.52, 0.51, 0.5, 0.49, 0.48, 0.475, 
and 0.47 (from left to right).
}
\label{fig:3}
\end{figure}

\begin{figure}
\epsfysize=\mysize
\begin{center}
\epsfbox{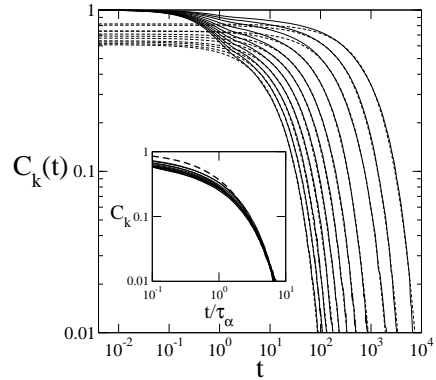}
\end{center}
\caption{
Log-log plot of $C_k(t)$ where $k=\pi$ and $p=0.56$, 0.55, 0.54, 0.53, 0.52, 
0.51, 0.5, 0.49, 0.48, 0.475, and 0.47 (from left to right). Dashed curves 
are fitting functions corresponding to the stretched exponential functions. 
Inset: time-driving force superposition principle. The dashed curve is a 
stretched exponential function with $\beta = 0.8$.
}
\label{fig:4}
\end{figure}

We find a two-step relaxation decay in the steady-state regime. We can see in  
Fig. \ref{fig:2} that the time interval of the first and second relaxation
step depends on the wave number $k$. Nevertheless, the form of the second 
relaxation step does not depend on $k$. For small enough $k$ we only have one 
step relaxation process. So, there is a wave number $k_e$ where the two-step
relaxation decay is lost for $k<k_e$. Fig. \ref{fig:3} shows $C_k(t)$ 
for $k=\pi$ and different values of the driving force $p$. The two-step
relaxation decay is observed from the highest value of $p=0.95$, but the 
time interval of the second step increases when $p$ is
decreased, i.e. when the system approaches to the criticality. This behavior is also 
found in other glassy systems, where the  time interval of the second relaxation step 
increases when the systems approach to the critical temperature.
As we can see in Fig. \ref{fig:4}, the second relaxation step can be fitted by a 
stretched exponential relaxation function Eq. (\ref{eq:1}) with the exponent 
$\beta=0.805\pm0.05$. This exponent is in practice independent of $p$ and it 
brings to the master equation
\be
C_k(t)=\widetilde{C_k}(t/\tau_\alpha)
\label{eq:2}
\ee
for $t> \tau_\alpha$, where $\widetilde{C_k}(t/\tau_\alpha)$ does not depend on $p$.
In glassy systems Eq. (\ref{eq:2}) is also called time-temperature superposition 
principle \cite{G91}, because the temperature plays the role of the driving force in
that systems. In the inset of Fig. \ref{fig:4} we show an equivalent time-driving 
force superposition principle for our system. This stretched exponential relaxation 
means that in the system there is a broad distribution of relaxation times 
\cite{RB94}.

\vglue 0.5cm

\begin{figure}
\epsfysize=\mysize
\begin{center}
\epsfbox{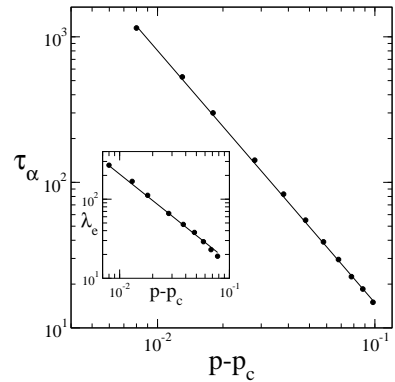}
\end{center}
\caption{
Log-log plot of the relaxation time $\tau_\alpha$, obtained by the stretched 
exponential fit in Fig. \ref{fig:4}, as a function of $p-p_c$,
for $p=0.56$, 0.55, 0.54, 0.53, 0.52, 0.51, 0.5, 0.49, 0.48, 0.475, 
and 0.47. The solid curve is a power law function 
$\tau_\alpha = 0.275(p-0.462)^{-1.733}$.
Inset: log-log plot of $\lambda_e$ as a function of $p-p_c$, for 
$p=0.54$, 0.53, 0.52, 0.51, 0.5, 0.49, 0.48, 0.475, and 0.47. The solid curve
is a power law function $\lambda_e=1.25(p-0.46)^{-1.1}$.
}
\label{fig:5}
\end{figure}

The relaxation time $\tau_\alpha$ can be obtained from the fit of $C_k(t)$ with a
stretched exponential function, it is shown in Fig. \ref{fig:5}. We see that it 
is very well fitted by a power law $\tau_\alpha \propto (p-p_c)^{-\nu_\parallel}$
where $p_c=0.462\pm 0.001$ and $\nu_\parallel=1.733\pm 0.001$. This means that
$\tau_\alpha$ is proportional to the characteristic distance of the clusters of 
pinning cells in the direction parallel to the interface $\xi_\parallel$, which
also diverges as $\tau_\alpha \sim \xi_\parallel \sim (p-p_c)^{-\nu_\parallel}$.

\vglue 0.5cm

We can obtain a wave length $\lambda_e=\pi/k_e$ in the direction perpendicular 
to the interface from the wave number $k_e$ where the two-step relaxation is lost.
In the inset of Fig. \ref{fig:5} we show $\lambda_e$ as a function of $(p-p_c)$. 
This length diverges as a power law $\lambda_e \propto (p-p_c)^{-\nu_\perp}$ with
$p_c=0.46\pm 0.01$ and $\nu_\perp=1.1\pm 0.01$. So that, $\lambda_e$ is 
proportional to the characteristic distance of the clusters of pinning cells 
in the direction perpendicular to the interface, 
$\lambda_e \sim \xi_\perp \sim (p-p_c)^{-\nu_\perp}$.

\vglue 0.5cm

We see that the characteristic length of the clusters of pinning cells in the 
direction parallel to the interface $\xi_\parallel$ sets the time scale of the 
stretched relaxation function $\widetilde{C_k}(t/\xi_\parallel)$. On the other 
hand, the stretched relaxation step is lost for $\lambda \ge \lambda_e$, that is 
for $\lambda \gtrsim \xi_\perp$. So, the stretched exponential relaxation is 
caused by the clusters of pinning cells. 

\section{Conclusions}

We have studied relaxation properties for growing interfaces in quenched
disordered media. We have used the TL model in which properties
of clusters of pinning cells are known. We have studied the relaxation
properties of the Fourier transform of the autocorrelation of the surface
height and found a two-step relaxation process in which the second 
step is well fitted by a stretched relaxation function with 
$\beta = 0.805\pm 005$. The relaxation time diverges as a power law and
it is proportional to the characteristic distance of the clusters of pinning 
cells in the direction parallel to the interface. The form of the second step 
relaxation does not depend on the wave number of the Fourier transform. This 
step is lost for a given wave length $\lambda_e$ which is proportional to the
characteristic distance of the clusters of pinning cells in the 
direction perpendicular to the interface. From these results, we can say that
the stretched exponential relaxation behavior is caused by the clusters of 
pinning cells, which is a direct consequence of the quenched noise as it 
happens in other glassy systems \cite{FC97}.

\acknowledgements

This work was supported in part by the project No. PI-60/00858/FS/01
from the Fundaci\'on S\'eneca, Regi\'on de Murcia.

\end{document}